\documentclass{elsart}
\usepackage{amsmath}
\usepackage{amsfonts}
\usepackage{amssymb}
\usepackage{graphicx}

\input{tcilatex}
\begin{document}

\begin{frontmatter}



\title{From Physics to Economics: An Econometric Example Using Maximum Relative Entropy}


\author{Adom Giffin}

\address{Princeton Institute for the Scientific and Technology of Materials\\ Princeton University\\ Princeton, NJ 08540,USA}
\begin{abstract}
Econophysics, is based on the premise that some ideas and methods from physics can be applied to economic situations. 
We intend to show in this paper how a physics concept such as entropy can be applied to an economic problem.
In so doing, we demonstrate how information in the form of observable data and moment
constraints are introduced into the method of Maximum relative Entropy (MrE).
A general example of updating with data and moments is shown. Two specific
econometric examples are solved in detail which can then be used as
templates for real world problems. A numerical example is compared to a
large deviation solution which illustrates some of the advantages of the MrE
method.
\end{abstract}

\end{frontmatter}

\section{Introduction}

\label{intro}

Methods of inference are not new to econometrics. In fact, one could say
that the subject is founded on inference methods. Econophysics, on the other
hand, is a much newer idea. It is based on the premise that some ideas and
methods from physics can be applied to economic situations. In this paper we
aim to show how a physics concept such as entropy can be applied to an
economic problem.

In 1957, Jaynes \cite{Jaynes57} showed that maximizing statistical mechanic
entropy for the purpose of revealing how gas molecules were distributed was
simply the maximizing of Shannon's information entropy \cite{Shannon1948}
with statistical mechanical information. The method was true for assigning
probabilities regardless of the information specifics. This idea lead to
MaxEnt or his use of the Method of Maximum Entropy for assigning
probabilities. This method has evolved to a more general method, the method
of Maximum (relative) Entropy (MrE) \cite%
{ShoreJohnson80,Skilling88,CatichaGiffin06} which has the advantage of not
only assigning probabilities but \emph{updating} them when new information
is given in the form of constraints on the family of allowed posteriors. One
of the draw backs of the MaxEnt method was the inability to include data.
When data was present, one used Bayesian methods. The methods were combined
in such a way that MaxEnt was used for assigning a prior for Bayesian
methods, as Bayesian methods could not deal with information in the form of
constraints, such as expected values. The main purpose of this paper is to
show both general and specific examples of how the MrE method can be applied
using data and moments\footnote{%
The constraints that we will be dealing with are more general than moments,
they are actually expected values. For simplicity we will refer to these
expected values as \emph{moments.}}.

The numerical example in this paper addresses a recent paper by Grendar and
Judge (GJ) \cite{EMME} where they consider the problem of criterion choice
in the context of large deviations (LD). Specifically, they attempt to
justify the method by Owen \cite{Owen2001} in a LD context with a new method
of their own. They support this idea by citing a paper in the econometric
literature by Kitamura and Stutzer \cite{KitamuraStutzer2002} who also use
LD\ to justify a particular empirical estimator. We attempt to simplify
their (GJ) initial problem by providing an example that is a bit more
practical. However, our example has the same issue; what does one do when
one has information in the form of an "average" of a large data set \emph{and%
} a small sample of that data set? We will show by example that the LD
approach is a special case of our method, the method of Maximum (relative)
Entropy

In section 2 we show a general example of updating simultaneously with two
different forms of information: moments and data. The solution resembles
Bayes' Rule. In fact, if there are no moment constraints then the method
produces Bayes rule \emph{exactly} \cite{GiffinCaticha07}. If there is no
data, then the MaxEnt solution is produced. The realization that MrE
includes not just MaxEnt but also Bayes' rule as special cases is highly
significant. It implies that MrE is \emph{capable of producing every aspect
of orthodox Bayesian inference} and proves the complete compatibility of
Bayesian and entropy methods. Further, it opens the door to tackling
problems that could not be addressed by either the MaxEnt or orthodox
Bayesian methods individually; problems in which one has data and moment
constraints.

In section 3 we comment on the problem of non-commuting constraints. We
discuss the question of whether they should be processed simultaneously, or
sequentially, and in what order. Our general conclusion is that these
different alternatives correspond to different states of information and
accordingly we expect that they will lead to different inferences.

In section 4, we provide two toy examples that illustrate potential economic
problems similar to the ones discussed in GJ. The two examples (ill-behaved
as mentioned in GJ) are solved in detail. The first example will demonstrate
how data and moments can be processed sequentially. This example is
typically how Bayesian statistics traditionally uses MaxEnt principles where
MaxEnt is used to create a prior for the Bayesian formulation. The second
example illustrates a problem that Bayes and MaxEnt alone cannot handle:
simultaneous processing of data and moments. These two examples will seem
trivially different but this is deceiving. They actually ask and answer two
completely different questions. It is this 'triviality' that is often a
source of confusion in Bayesian literature and therefore we wish to expose
it.

In section 6 we compare a numerical example that is solved by MrE and one
that is solved by GJ's method. Since GJ's solution comes out of LD, they
rely on asymptotic arguments; one assumes an infinite sample set which is
not necessarily realistic. The MrE method does not need such assumptions to
work and therefore can process finite amounts of data well. However, when
MrE is taken to asymptotic limits one recovers the same solutions that the
large deviation methods produce.

\section{Simultaneous updating with moments and data}

\label{sec:2}

Our first concern when using the MrE method to update from a prior to a
posterior distribution is to define the space in which the search for the
posterior will be conducted. We wish to infer something about the values of
one or several quantities, $\theta \in \Theta $, on the basis of three
pieces of information: prior information about $\theta $ (the prior), the
known relationship between $x$ \emph{and} $\theta $ (the model), and the
observed values of the data $x\in \mathcal{X}$. Since we are concerned with
both $x$ \emph{and} $\theta $, the relevant space is neither $\mathcal{X}$
nor $\Theta $ but the product $\mathcal{X}\times \Theta $ and our attention
must be focused on the joint distribution $P\left( x,\theta \right) $. The
selected joint posterior $P_{\text{new}}(x,\theta )$ is that which maximizes
the entropy\footnote{%
In the MrE terminology, we "maximize" the negative relative entropy, $S$ so
that $S\leq 0.$ This is the same as minimizing the relative entropy.},%
\begin{equation}
S[P,P_{\text{old}}]=-\int dxd\theta ~P\left( x,\theta \right) \log \frac{%
P\left( x,\theta \right) }{P_{\text{old}}\left( x,\theta \right) }~,
\label{entropy}
\end{equation}%
subject to the appropriate constraints. $P_{\text{old}}\left( x,\theta
\right) $ contains our prior information which we call the \emph{joint prior}%
. To be explicit,%
\begin{equation}
P_{\text{old}}\left( x,\theta \right) =P_{\text{old}}\left( \theta \right)
P_{\text{old}}\left( x|\theta \right) ~,  \label{joint prior}
\end{equation}%
where $P_{\text{old}}\left( \theta \right) $ is the traditional Bayesian
prior and $P_{\text{old}}\left( x|\theta \right) $ is the likelihood. It is
important to note that they \emph{both} contain prior information. The
Bayesian prior is defined as containing prior information. However, the
likelihood is not traditionally thought of in terms of prior information. Of
course it is reasonable to see it as such because the likelihood represents
the model (the relationship between $\theta $ and $x)$ that has already been
established. Thus we consider both pieces, the Bayesian prior and the
likelihood to be \emph{prior} information.

The new information is the \emph{observed data}, $x^{\prime }$, which in the
MrE framework must be expressed in the form of a constraint on the allowed
posteriors. The family of posteriors that reflects the fact that $x$ is now
known to be $x^{\prime }$ is such that%
\begin{equation}
P(x)=\int d\theta ~P\left( x,\theta \right) =\delta \left( x-x^{\prime
}\right) ~,  \label{data}
\end{equation}%
where $\delta \left( x-x^{\prime }\right) $ is the Dirac delta function.
This amounts to an \emph{infinite} number of constraints: there is one
constraint on $P\left( x,\theta \right) $ for each value of the variable $x$
and each constraint will require its own Lagrange multiplier $\lambda (x)$.
Furthermore, we impose the usual normalization constraint, 
\begin{equation}
\int dxd\theta ~P\left( x,\theta \right) =1~,  \label{Normalization}
\end{equation}%
and include additional information about $\theta $ in the form of a
constraint on the expected value of some function $f\left( \theta \right) $, 
\begin{equation}
\int dxd\theta \,P\left( x,\theta \right) f\left( \theta \right)
=\left\langle f\left( \theta \right) \right\rangle =F~.  \label{moment}
\end{equation}%
Note: an additional constraint in the form of $\int dxd\theta P(x,\theta
)g(x)=\left\langle g\right\rangle =G$ could only be used when it does not
contradict the data constraint (\ref{data}). Therefore, it is redundant and
the constraint would simply get absorbed when solving for $\lambda (x)$. We
also emphasize that constraints imposed at the level of the prior need not
be satisfied by the posterior. What we do here differs from the standard
Bayesian practice in that we \emph{require} the constraint to be satisfied
by the posterior distribution.

We proceed by maximizing (\ref{entropy}) subject to the above constraints.
The purpose of maximizing the entropy is to determine the value for $P$ when 
$S=0.$ Meaning, we want the value of $P$ that is closest to $P_{\text{old}}$
given the constraints. The calculus of variations is used to do this by
varying $P\rightarrow \delta P$, i.e. setting the derivative with respect to 
$P$ equal to zero. The Lagrange multipliers $\alpha ,\beta $ and $\lambda
(x) $ are used so that the $P$ that is chosen satisfies the constraint
equations. The actual values are determined by the value of the constraints
themselves. We now provide the detailed steps in this maximization process.

First we setup the variational form with the Lagrange multipiers, 
\begin{equation}
\delta P\left( x,\theta \right) \left\{ 
\begin{array}{c}
S[P,P_{\text{old}}]+\alpha \left[ \int dxd\theta P\left( x,\theta \right) -1%
\right] \\ 
+\beta \left[ \int dxd\theta P\left( x,\theta \right) f\left( \theta \right)
-F\right] \\ 
+\int dx\lambda (x)\left[ \int d\theta P\left( x,\theta \right) -\delta
\left( x-x%
{\acute{}}%
\right) \right]%
\end{array}%
\right\} =0~.
\end{equation}%
We expand the entropy function (\ref{entropy}), 
\begin{equation}
\delta P\left( x,\theta \right) \left\{ 
\begin{array}{c}
-\int dxd\theta ~P\left( x,\theta \right) \log P\left( x,\theta \right) \\ 
+\int dxd\theta ~P\left( x,\theta \right) \log P_{\text{old}}\left( x,\theta
\right) \\ 
+\alpha \left[ \int dxd\theta P\left( x,\theta \right) -1\right] \\ 
+\beta \left[ \int dxd\theta P\left( x,\theta \right) f\left( \theta \right)
-F\right] \\ 
+\int dx\lambda (x)\left[ \int d\theta P\left( x,\theta \right) -\delta
\left( x-x%
{\acute{}}%
\right) \right]%
\end{array}%
\right\} =0~.
\end{equation}%
Next, vary the functions with respect to $P\left( x,\theta \right) ,$ 
\begin{equation}
\left\{ 
\begin{array}{c}
-\int dxd\theta ~\delta P\left( x,\theta \right) \log P\left( x,\theta
\right) -\int dxd\theta ~P\left( x,\theta \right) \frac{1}{P\left( x,\theta
\right) }\delta P\left( x,\theta \right) \\ 
+\int dxd\theta ~\delta P\left( x,\theta \right) \log P_{\text{old}}\left(
x,\theta \right) +0 \\ 
+\alpha \left[ \int dxd\theta ~\delta P\left( x,\theta \right) \right] \\ 
+\beta \left[ \int dxd\theta ~\delta P\left( x,\theta \right) f\left( \theta
\right) \right] \\ 
+\int dx\lambda (x)\left[ \int d\theta ~\delta P\left( x,\theta \right) %
\right]%
\end{array}%
\right\} =0~,
\end{equation}%
which can be rewritten as 
\begin{equation*}
\int dxd\theta \left\{ -\log P\left( x,\theta \right) -1+\log P_{\text{old}%
}\left( x,\theta \right) +\alpha +\beta f\left( \theta \right) +\lambda
(x)\right\} \delta P\left( x,\theta \right) =0~.
\end{equation*}%
The terms inside the brackets must sum to zero, therefore we can write, 
\begin{equation}
\log P\left( x,\theta \right) =\log P_{\text{old}}\left( x,\theta \right)
-1+\alpha +\beta f\left( \theta \right) +\lambda (x)~
\end{equation}%
or 
\begin{equation}
P_{\text{new}}\left( x,\theta \right) =P_{\text{old}}\left( x,\theta \right)
e^{\left( -1+\alpha +\beta f\left( \theta \right) +\lambda (x)\right) }~
\label{Posterior 1}
\end{equation}

In order to determine the Lagrange multipliers, we substitute our solution (%
\ref{Posterior 1}) into the various constraint equations. The constant $%
\alpha $ is eleiminated by substituting (\ref{Posterior 1}) into (\ref%
{Normalization}), 
\begin{equation}
\int dxd\theta ~P_{\text{old}}\left( x,\theta \right) e^{\left( -1+\alpha
+\beta f\left( \theta \right) +\lambda (x)\right) }=1~.
\end{equation}%
Dividing both sides by the constant $e^{\left( -1+\alpha \right) }$, 
\begin{equation}
\int dxd\theta ~P_{\text{old}}\left( x,\theta \right) e^{\beta f\left(
\theta \right) +\lambda (x)}=e^{\left( 1-\alpha \right) }~.
\end{equation}%
Then substituting back into (\ref{Posterior 1}) yields 
\begin{equation}
P_{\text{new}}(x,\theta )=P_{\text{old}}\left( x,\theta \right) \frac{%
e^{\lambda (x)+\beta f\left( \theta \right) }}{Z}~,  \label{Posterior 2}
\end{equation}%
where 
\begin{equation}
Z=e^{1-\alpha }=\int dxd\theta e^{\beta f\left( \theta \right) +\lambda
(x)}P_{\text{old}}\left( x,\theta \right) ~.
\end{equation}%
\newline
In the same fashion, the Lagrange multipliers $\lambda (x)$ are determined
by substituting (\ref{Posterior 2}) into (\ref{data})%
\begin{equation}
\int d\theta ~P_{\text{old}}\left( x,\theta \right) \frac{e^{\lambda
(x)+\beta f\left( \theta \right) }}{Z}=\delta \left( x-x^{\prime }\right) ~
\end{equation}%
or%
\begin{equation}
e^{\lambda (x)}=\frac{Z}{\int d\theta e^{\beta f\left( \theta \right) }P_{%
\text{old}}\left( x,\theta \right) }\delta (x-x%
{\acute{}}%
)~.
\end{equation}%
The posterior now becomes%
\begin{equation}
P_{\text{new}}(x,\theta )=P_{\text{old}}\left( x,\theta \right) \delta (x-x%
{\acute{}}%
)\frac{e^{\beta f\left( \theta \right) }}{\zeta (x,\beta )}~,
\label{Posterior-Both}
\end{equation}%
where $\zeta (x,\beta )=\int d\theta e^{\beta f\left( \theta \right) }P_{%
\text{old}}\left( x,\theta \right) .$

The Lagrange multiplier $\beta $ is determined by first substituting (\ref%
{Posterior-Both}) into (\ref{moment}),%
\begin{equation}
\int dxd\theta \left[ P_{\text{old}}\left( x,\theta \right) \delta (x-x%
{\acute{}}%
)\frac{e^{\beta f\left( \theta \right) }}{\zeta (x,\beta )}\right] f\left(
\theta \right) =F~.
\end{equation}%
Integrating over $x$ yields,%
\begin{equation}
\frac{\int d\theta e^{\beta f\left( \theta \right) }P_{\text{old}}(x^{\prime
},\theta )f\left( \theta \right) }{\zeta (x^{\prime },\beta )}=F~,
\label{F1}
\end{equation}%
where $\zeta (x,\beta )\rightarrow \zeta (x^{\prime },\beta )=\int d\theta
e^{\beta f\left( \theta \right) }P_{\text{old}}(x^{\prime },\theta )$. Now $%
\beta $ can be determined rewriting (\ref{F1}) as 
\begin{equation}
\frac{\partial \ln \zeta (x^{\prime },\beta )}{\partial \beta }=F~.
\label{F}
\end{equation}

The final step is to marginalize the posterior, $P_{\text{new}}(x,\theta )$
over $x$ to get our updated probability,%
\begin{equation}
P_{\text{new}}(\theta )=P_{\text{old}}(x^{\prime },\theta )\frac{e^{\beta
f\left( \theta \right) }}{\zeta (x^{\prime },\beta )}
\end{equation}%
Additionally, this result can be rewritten using the product rule ($P\left(
x,\theta \right) =P\left( x\right) P\left( \theta |x\right) $) as%
\begin{equation}
P_{\text{new}}(\theta )=P_{\text{old}}\left( \theta \right) P_{\text{old}%
}(x^{\prime }|\theta )\frac{e^{\beta f\left( \theta \right) }}{\zeta
^{\prime }\left( x^{\prime },\beta \right) }~,
\end{equation}%
where $\zeta ^{\prime }(x^{\prime },\beta )=\int d\theta e^{\beta f\left(
\theta \right) }P_{\text{old}}\left( \theta \right) P_{\text{old}}(x^{\prime
}|\theta ).$ The right side resembles Bayes theorem, where the term $P_{%
\text{old}}(x^{\prime }|\theta )$ is the standard Bayesian likelihood and $%
P_{\text{old}}\left( \theta \right) $ is the prior. The exponential term is
a \emph{modification} to these two terms. In an effort to put some names to
these pieces we will call the standard Bayesian likelihood the \emph{%
likelihood} and the exponential part the \emph{likelihood modifier} so that
the product of the two gives the \emph{modified likelihood}. The denominator
is the normalization or \emph{marginal modified likelihood}. Notice when $%
\beta =0$ (no moment constraint) we recover Bayes' rule. For $\beta \neq 0$
Bayes' rule is modified by a \textquotedblleft canonical\textquotedblright\
exponential factor.

\section{Commutivity of constraints}

\label{sec:3}

When we are confronted with several constraints, such as in the previous
section, we must be particularly cautious. In what order should they be
processed? Or should they be processed at the same time? The answer depends
on the nature of the constraints and the question being asked \cite%
{GiffinCaticha07}.

We refer to constraints as \emph{commuting} when it makes no difference
whether they are processed simultaneously or sequentially. The most common
example of commuting constraints is Bayesian updating on the basis of data
collected in multiple experiments. For the purpose of inferring $\theta $ it
is well known that the order in which the observed data $x^{\prime
}=\{x_{1}^{\prime },x_{2}^{\prime },\ldots \}$ is processed does not matter.
The proof that MrE is completely compatible with Bayes' rule implies that
data constraints implemented through $\delta $ functions, as in (\ref{data}%
), commute just as they do in Bayes.

It is important to note that when an experiment is repeated it is common to
refer to the value of $x$ in the first experiment and the value of $x$ in
the second experiment. This is a dangerous practice because it obscures the
fact that we are actually talking about \emph{two} separate variables. We do
not deal with a single $x$ but with a composite $x=(x_{1},x_{2})$ and the
relevant space is $\mathcal{X}_{1}\times \mathcal{X}_{2}\times \Theta $.
After the first experiment yields the value $x_{1}^{\prime }$, represented
by the constraint $c_{1}:P(x_{1})=\delta \left( x_{1}-x_{1}^{\prime }\right) 
$, we can perform a second experiment that yields $x_{2}^{\prime }$ and is
represented by a second constraint $c_{2}:P(x_{2})=\delta \left(
x_{2}-x_{2}^{\prime }\right) $. These constraints $c_{1}$ and $c_{2}$
commute because they refer to \emph{different} variables $x_{1}$ and $x_{2}$

As a side note, use of a $\delta $ function has been criticized in that by
implementing it, the probability is completely constrained, thus it cannot
be updated by future information. This is certainly true! An experiment,
once performed and its outcome observed, cannot be \emph{un-performed} and
its result cannot be \emph{un-observed} by subsequent experiments. Thus,
imposing one constraint does not imply a revision of the other.

\begin{figure}[!t]
\center
\resizebox{.5\columnwidth}{!}
{\includegraphics[draft=false]{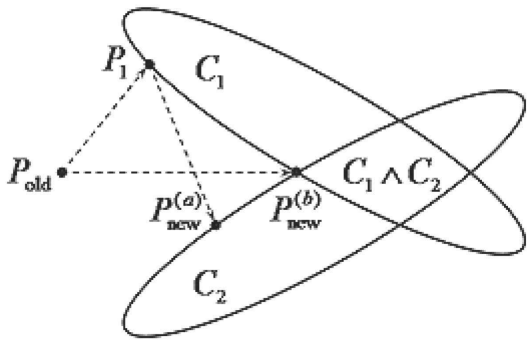}}
\caption{Illustrating the difference between processing two constraints $%
C_{1}$ and $C_{2}$ sequentially ($P_{\text{old}}\rightarrow P_{1}\rightarrow
P_{\text{new}}^{(a)}$) and simultaneously ($P_{\text{old}}\rightarrow P_{%
\text{new}}^{(b)}$ or $P_{\text{old}}\rightarrow P_{1}\rightarrow P_{\text{%
new}}^{(b)}$).}
\label{Fig1:a}
\end{figure}

In general constraints need not commute and when this is the case the order
in which they are processed is critical. For example, suppose the prior is $%
P_{\text{old}}$ and we receive information in the form of a constraint, $%
C_{1}$. To update we maximize the entropy $S[P,P_{\text{old}}]$ subject to $%
C_{1}$ leading to the posterior $P_{1}$ as shown in Fig 1. Next we receive a
second piece of information described by the constraint $C_{2}$. At this
point we can proceed in essentially two different ways:

a) Sequential updating -

Having processed $C_{1}$, we use $P_{1}$ as the current prior and maximize $%
S[P,P_{1}]$ subject to the new constraint $C_{2}$. This leads us to the
posterior $P_{\text{new}}^{(a)}$.

b) Simultaneous updating -

Use the original prior $P_{\text{old}}$ and maximize $S[P,P_{\text{old}}]$
subject to both constraints $C_{1}$ and $C_{2}$ simultaneously. This leads
to the posterior $P_{\text{new}}^{(b)}$. At first sight it might appear that
there exists a third possibility of\ simultaneous updating: (c) use $P_{1}$
as the current prior and maximize $S[P,P_{1}]$ subject to both constraints $%
C_{1}$ and $C_{2}$ simultaneously. Fortunately, and this is a valuable check
for the consistency of the ME method, it is easy to show that case (c) is
equivalent to case (b). Whether we update from $P_{\text{old}}$ or from $%
P_{1}$ the selected posterior is $P_{\text{new}}^{(b)}$.

To decide which path (a) or (b) is appropriate, we must be clear about how
the MrE method treats constraints. The MrE machinery interprets a constraint
such as $C_{1}$ in a very mechanical way: all distributions satisfying $%
C_{1} $ are in principle allowed and all distributions violating $C_{1}$ are
ruled out.

Updating to a posterior $P_{1}$ consists precisely in revising those aspects
of the prior $P_{\text{old}}$ that disagree with the new constraint $C_{1}$.
However, there is nothing final about the distribution $P_{1}$. It is just
the best we can do in our current state of knowledge and we fully expect
that future information may require us to revise it further. Indeed, when
new information $C_{2}$ is received we must reconsider whether the original $%
C_{1}$ remains valid or not. Are \emph{all} distributions satisfying the new 
$C_{2}$ really allowed, even those that violate $C_{1}$? If this is the case
then the new $C_{2}$ takes over and we update from $P_{1}$ to $P_{\text{new}%
}^{(a)}$. The constraint $C_{1}$ may still retain some lingering effect on
the posterior $P_{\text{new}}^{(a)}$ through $P_{1},$ but in general $C_{1}$
has now become obsolete.

Alternatively, we may decide that the old constraint $C_{1}$ retains its
validity. The new $C_{2}$ is not meant to revise $C_{1}$ but to provide an
additional refinement of the family of allowed posteriors. In this case the
constraint that correctly reflects the new information is not $C_{2}$ but
the more restrictive space where $C_{1}$ and $C_{2}$ overlap. The two
constraints should be processed simultaneously to arrive at the correct
posterior $P_{\text{new}}^{(b)}$.

To summarize: sequential updating is appropriate when old constraints become
obsolete and are superseded by new information; simultaneous updating is
appropriate when old constraints remain valid. The two cases refer to
different states of information and therefore \emph{we expect} that they
will result in different inferences. These comments are meant to underscore
the importance of understanding what information is being processed; failure
to do so will lead to errors that do not reflect a shortcoming of the MrE
method but rather a misapplication of it.

\section{An econometric problem: sequential updating}

\label{sec:4}

This is an example of a problem using the MrE method:. The general
background information is that a factory makes $k$ different kinds of bouncy
balls. For reference, they assign each different kind with a number, $%
f_{1},f_{2},...f_{k}$. They ship large boxes of them out to stores.
Unfortunately, there is no mechanism that regulates how many of each ball
goes into the boxes, therefore we do not know the amount of each kind of
ball in any of the boxes.

For this problem we are informed that the company does know the average of
all the kinds of balls, $F$ that is produced by the factory over the time
that they have been in existence. This is information about the \emph{factory%
}. By using this information with MrE we get what one would get with the old
MaxEnt method, a distribution of balls for the whole factory.

However, we would like to know the probability of getting a certain kind of
ball in a particular box. Therefore, we are allowed to randomly select a few
balls, $n$ from the particular box in question and count how many of each
kind we get, $m_{1},m_{2}...m_{k}$ (or perhaps we simply open the box and
look at the balls on the surface). This is information about the \emph{%
particular box}. Now let us put the above example in a more mathematical
format.

Let the set of possible outcomes be represented by, $\Omega
=\{f_{1},f_{2},...f_{k}\}$ from a sample where the total number of balls, $%
N\rightarrow \infty $\footnote{%
It is not necessary for $N\rightarrow \infty $ for the ME method to work. We
simply wish to use the description of the problem that is common in
information-theoretic examples. It must be strongly noted however that in
general a sample average is not an expectation value.} and whose sample
average is $F.$ Further, let us draw a \emph{data} sample of size $n,$ from
a particular subset of the original sample, $\omega $ where $\omega \in
\Omega $ and whose outcomes are counted and represented as $%
m=(m_{1},m_{2}...m_{k})$ where $n=\sum\nolimits_{i}^{k}m_{i}$. We would like
to determine the probability of getting \emph{any} particular type in one
draw ($\theta =\{\theta _{1},\theta _{2}...\theta _{k}\}$) out of the subset
given the information. To do this we start with the appropriate joint
entropy,%
\begin{equation}
S[P,P_{\text{old}}]~\text{=}-\sum\limits_{m}\int d\theta P(m,\theta |n)\log 
\frac{P\left( m,\theta |n\right) }{P_{\text{old}}(m,\theta |n)}.
\label{entropy P1}
\end{equation}%
We then maximize this entropy with respect to $P(m,\theta |n)$ to process
the first piece of information that we have which is the moment constraint, $%
C_{1}$ that is related to the factory,%
\begin{equation}
C_{1}:\left\langle f\left( \theta \right) \right\rangle =F\quad \text{where}%
\quad f\left( \theta \right) =\sum\nolimits_{i}^{k}f_{i}\theta _{i}~,
\end{equation}%
subject to normalization, where $\theta =\{\theta _{1},\theta _{2}...\theta
_{k}\},$ $m=\left( m_{1}...m_{k}\right) $ and where\footnote{%
The use of the $\delta $ function in both (18)\ and (19) are used to clarify
the summation notaion used in (16). They are not \emph{information}
constraints.on $P$ as in (18) and later (25).}%
\begin{equation}
\sum\limits_{m}=\sum\limits_{m_{1}\ldots m_{k}=0}^{n}\delta \left(
\sum\nolimits_{i=1}^{k}m_{i}-n\right) ~,
\end{equation}%
and%
\begin{equation}
\int d\theta =\int d\theta _{1}\ldots d\theta _{k}\,\delta \left(
\sum\nolimits_{i=1}^{k}\theta _{i}-1\right) ~.
\end{equation}%
This yields,%
\begin{equation}
P_{\text{1}}(m,\theta |n)=P_{\text{old}}(m,\theta |n)\frac{e^{\lambda
f\left( \theta \right) }}{Z_{1}}~,  \label{joint P1}
\end{equation}%
where the normalization constant $Z_{1}$ and the Lagrange multiplier $%
\lambda $ are determined from%
\begin{equation}
Z_{1}=\int d\theta \,e^{\lambda f\left( \theta \right) }P_{\text{old}%
}(\theta |n)\quad \text{and}\quad \frac{\partial \log Z_{1}}{\partial
\lambda }=F~.  \label{Z1}
\end{equation}%
We need to determine what to use for our joint prior,%
\begin{equation}
P_{\text{old}}(m,\theta |n)=P_{\text{old}}(m^{\prime }|\theta ,n)P_{\text{old%
}}(\theta |n)
\end{equation}%
in our problem. The mathematical representation of the\ situation where we
wish to know the probability of selecting $m_{i}$ balls of the $i^{th}$ type
from a sample of $n$ balls of $k$-types is simply the multinomial
distribution. Therefore, the equation that we will use for our model, the 
\emph{likelihood,} $P_{\text{old}}(m^{\prime }|\theta ,n)$ is,%
\begin{equation}
P_{\text{old}}(m_{1}...m_{k}|\theta _{1}...\theta _{k},n)=\frac{n!}{%
m_{1}!...m_{k}!}\theta _{1}^{m_{1}}..\theta _{k}^{m_{k}}~.
\end{equation}%
Since at this point we are completely ignorant of $\theta ,$ we use a \emph{%
prior} that is flat, thus $P_{\text{old}}(\theta |n)=$ \emph{constant}.
Being a constant, the prior can come out of the integral and cancels with
the same constant in the numerator. (Also, the particular form of $P_{\text{%
old}}(\theta |n)$ is not important for our current purpose so for the sake
of definiteness we can choose it flat for our example. There are most likely
better choices for priors, such as a Jeffrey's prior.) Thus, after
marginalizing over $m,$ the joint distribution (\ref{joint P1}) can be
rewritten as%
\begin{equation}
P_{1}(\theta )=\frac{e^{\lambda f\left( \theta \right) }}{Z_{1}}~.
\label{P1}
\end{equation}

Now we wish to process the next piece of information which is the data
constraint,%
\begin{equation}
C_{2}:P(m)=\delta _{mm^{\prime }}~.  \label{C2a}
\end{equation}%
Here we use a Kronecker delta function since $m$ is discrete in this
example. Our goal is to infer the $\theta $ that apply to our particular
box. The original constraint $C_{1}$ applies to the whole factory while the
new constraint $C_{2}$ refers to the actual box of interest and thus takes
precedence over $C_{1}.$ As $n\rightarrow \infty $ we expect $C_{1}$ to
become less and less relevant. Therefore the two constraints should be
processed sequentially.

We maximize again with our new information which yields,%
\begin{equation}
P_{\text{new}}^{(a)}(m,\theta )=\delta _{mm^{\prime }}P_{1}(\theta |m)~.
\end{equation}%
Marginalizing over $m$ and using (\ref{P1}) the final posterior for $\theta $
is%
\begin{equation}
P_{\text{new}}^{(a)}(\theta )=P_{1}(\theta |m^{\prime })=P_{\text{old}%
}(m^{\prime }|\theta )\frac{e^{\lambda f\left( \theta \right) }}{Z_{2}}~.
\label{posterior a}
\end{equation}%
where%
\begin{equation}
Z_{2}=\int d\theta \,e^{\lambda f\left( \theta \right) }P_{\text{old}%
}(m^{\prime }|\theta )~.  \label{Z2}
\end{equation}

Those familiar with using MaxEnt and Bayes will undoubtedly recognize that (%
\ref{posterior a}) is precisely the result obtained by using MaxEnt to
obtain a prior, in this case $P_{1}(\theta )$ given in (\ref{P1}), and then
using Bayes' rule to take the data into account. This familiar result has
been derived in detail for two reasons: first, to reassure the readers that
MrE does reproduce the standard solutions to standard problems and second,
to establish a contrast with the example discussed next. NOTE: Since the
constraints $C_{1}$ and $C_{2}$ do not commute one will get a different
result if they are processed in a different order.

\section{An econometric problem: simultaneous updating}

\label{sec:5}

This is another example of a problem using the MrE method:. The general
background information is the same as the previous example. For this problem
we are informed that the company knows the average of all the kinds of
balls, $F$ in \emph{each} \emph{box}. By using this information with MrE we
get what one would get with the old MaxEnt method, a distribution of balls
for each box.

However, we still would like to know the probability of getting a certain
kind of ball in a particular box and we are allowed to randomly select a few
balls, $n$ from the particular box in question once again. Since both of
these pieces of information apply to the \emph{same} box, they must be
processed simultaneously. In other words, both constraints must hold,
always. We proceed as in the first example by maximizing (\ref{entropy P1})
subject to normalization and the following constraints simultaneously,%
\begin{equation}
C_{3}:\left\langle f\left( \theta \right) \right\rangle =F\quad \text{where}%
\quad f\left( \theta \right) =\sum\nolimits_{i}^{k}f_{i}\theta _{i}~,
\end{equation}%
(notice $C_{3}\neq C_{1}$ because they are two difference pieces of
information) and%
\begin{equation}
C_{2}:P(m)=\delta _{mm^{\prime }}~.
\end{equation}%
This yields,%
\begin{equation}
P_{\text{new}}^{(b)}(\theta )=P_{1}(\theta |m^{\prime })=P_{\text{old}%
}(m^{\prime }|\theta )\frac{e^{\beta f\left( \theta \right) }}{\zeta }~,
\label{posterior b}
\end{equation}%
where%
\begin{equation}
\zeta =\int d\theta \,e^{\beta f\left( \theta \right) }P_{\text{old}%
}(m^{\prime }|\theta )~,  \label{zeta b}
\end{equation}%
and%
\begin{equation}
F=\frac{\partial \log \zeta }{\partial \beta }~.  \label{F b}
\end{equation}

This looks like the sequential case (\ref{posterior a}), but there is a
crucial difference: $\beta \neq \lambda $ and $\zeta \neq Z_{2}$. In the
sequential updating case, the multiplier $\lambda $ is chosen so that the
intermediate $P_{1}$ satisfies $C_{1}$ while the posterior $P_{\text{new}%
}^{(a)}$ only satisfies $C_{2}$. In the simultaneous updating case the
multiplier $\beta $ is chosen so that the posterior $P_{\text{new}}^{(b)}$
satisfies both $C_{1}$ and $C_{2}$ or $C_{1}\wedge C_{2}$. Ultimately, the
two distributions $P_{\text{new}}(\theta )$ are different because they refer
to different problems. For more examples using this method see \cite%
{GiffinCaticha07}.

\section{Numerical examples}

\label{Numerical}

The purpose of this section is two fold: First, we would like to provide a
numerical example of a MrE solution. Second, we wish to examine a current,
relevant econometric solution proposed by GJ in \cite{EMME} using the method
of types, specifically large deviation theory, for an "ill-posed" problem
that is similar to the one discussed in section 5. This solution will be
compared with a solution using MrE.

To summarize the problem once again: The factory makes $k$ different kinds
of bouncy balls and for reference, they assign each different type with a
number, $f_{1},f_{2},...f_{k}$. We are informed that the company knows the
expected type of ball, $F$ in each box over the time that they have been in
existence. We would like a better idea of how many balls are in each box so
we randomly select a few balls, $n$ from a particular box and count how many
of each type we get, $m_{1},m_{2}...m_{k}$.

Or stated in a more mathematical format: Let the set of possible outcomes of
a be represented by, $\Omega =\{f_{1},f_{2},...f_{k}\}$ from a sample where
the total number of balls, $N\rightarrow \infty $. and where the average of
the types of balls is $F.$ Further, let us draw a \emph{data} sample of size 
$n,$ from the original sample, whose outcomes are counted and represented as 
$m=(m_{1},m_{2}...m_{k})$ where $n=\sum\nolimits_{i}^{k}m_{i}$. The problem
becomes ill-posed when the sample average of the counts%
\begin{equation}
S_{Avg}=\frac{1}{n}\sum_{i}f_{i}m_{i}  \label{Sample Avg}
\end{equation}%
significantly deviates from the expected average of the types, $F.$

We would like to determine the probability of getting \emph{any} particular
outcome in one draw ($\theta =\{\theta _{1},\theta _{2}...\theta _{k}\}$)
given the information.

\subsection{Sanov's theorem solution}

In \cite{EMME} a form of Sanov's theorem is used. Here we give a brief
description of Sanov's theorem. It is not intended to be a proof or
exhaustive. It is simply shown to give a general indication of the basis for
the solution in \cite{EMME}. The key equation is (\ref{Econ Sanov Solution}%
). For a more detailed proof and explanation see \cite{CoverThomas}.

Sanov's theorem -

Let $X_{1}\ldots X_{n}$ be independent and identically distributed (i.i.d.)
with values in an arbitrary set $\chi $ with common distribution $Q(x)$. Let 
$E\subseteq \mathcal{P}$ be a set of probability distributions. Then,%
\begin{equation}
Q^{n}(E)=Q^{n}(E\cap \mathcal{P}_{n})\leq (n+1)^{\left\vert \mathcal{X}%
\right\vert }2^{-nD(P^{\ast }||Q)}~,
\end{equation}%
where 
\begin{equation}
P^{\ast }=\arg \min_{P\in E}D(P||Q)
\end{equation}%
is the distribution in $E$ that is closest to $Q$ in the relative entropy or
information divergence, 
\begin{equation}
D(P||Q)=\sum\limits_{x\in \chi }dxP(x)\log \frac{P\left( x\right) }{Q(x)}
\label{info divergence}
\end{equation}
and $n$ is the number of types. If in addition, the set $E$ is the closure
of its interior, 
\begin{equation}
\frac{1}{n}\log Q^{n}(E)\rightarrow -D(P^{\ast }||Q)~.
\end{equation}%
The two equations become equal in the asymptotic limit. Essentially what
this theorem says is that in the asymptotic limit, the frequency of the
sample $Q$, can be used to produce an estimate, $P^{\ast }$ of the "true"
probability, $\mathcal{P}$ by way of minimizing the relative entropy (\ref%
{info divergence}).

For our problem, the solution for the probability using Sanov is of the
form, 
\begin{equation}
P^{\ast }=\frac{Qe^{\eta _{i}f_{i}}}{\sum_{i}Qe^{\eta _{i}f_{i}}}~,
\label{Econ Sanov Solution}
\end{equation}%
where $Q$ for our problem is the frequency of the counts, $m/n$ and $\eta $
is a Lagrange multiplier that is determined by the sample average (\ref%
{Sample Avg}), not an expected value as in our method. This solution seems
very similar to our general solution using the MrE method (\ref{posterior b}%
) in which we also minimize an entropy (maximize our negative relative
entropy). We could even think of $Q$ as a kind of joint prior and
likelihood. However, there are many differences in the two methods, but the
most glaring is that the GJ solution is only valid in the asymptotic case.
We are not handicapped by this when MrE is used.

\subsection{Comparing the methods}

We illustrate the differences between the methods be examining a specific
version of the above problem: Let the there be three kinds of balls labeled
1, 2 and 3. So for this problem we have $f_{1}=1,$ $f_{2}=2$ and $f_{3}=3.$
Further, we are given information regarding the expected value of each box, $%
F.$ For our example this value will be, $F=2.3.$ Notice that this implies
that on the average there are more $3$'s in each box. Next we take a sample
of one of the boxes where $m_{1}^{\prime }=11,$ $m_{2}^{\prime }=2$ and $%
m_{3}^{\prime }=7.$

Using the MrE method in the same way that we have in each of the previous
sections, we arrive at a posterior solution after maximizing the proper
entropy subject to the constraints,%
\begin{equation}
P_{\text{MrE}}(\theta _{1},\theta _{2})=\frac{1}{\zeta _{\text{MrE}}}%
e^{\beta (-2\theta _{1}-\theta _{2}+3)}\theta _{1}^{11}\theta
_{2}^{2}(1-\theta _{1}-\theta _{2})^{7}~.  \label{MrE Posterior}
\end{equation}%
where the Lagrange multiplier $\beta $ was determined using Newton's method
on the equation (\ref{F b}) and found to be $\beta =14.1166.$ We show the
relationship between $\beta $ and $F$ in Fig 2.

\begin{figure}[!t]
\center
\rotatebox {-90}{\resizebox{2.5in}{4.5in} 
{\includegraphics[draft=false]{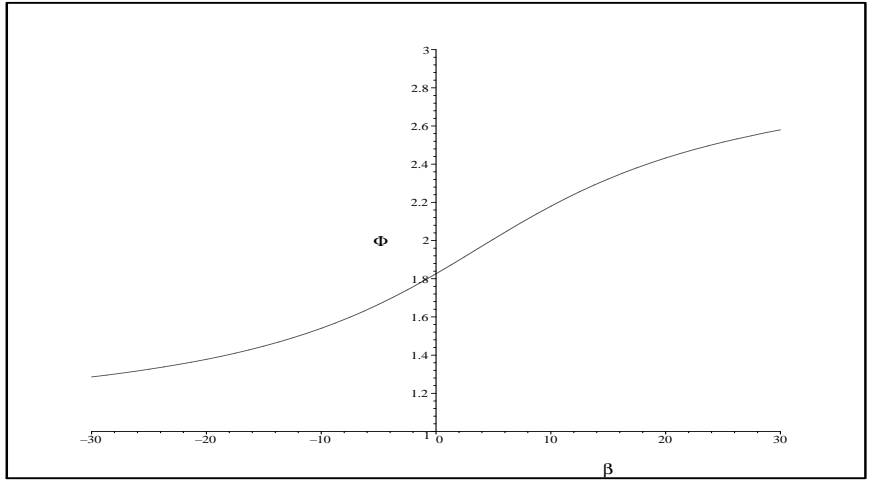}}}
\caption{This figure shows the relationship between $\protect\beta $ and $%
\Phi (\Phi =F(\protect\beta )).$ Notice that as the value for $\Phi $
approaches the extremities of the outcomes, $\protect\beta $ approaches
infinity.}
\label{Fig2:a}
\end{figure}

This result is then put into our calculation of $\zeta _{\text{MrE}}$ so
that $\zeta _{\text{MrE}}=1874.1247.$ Two plots are provided that show the
marginal distributions of $\theta _{1}$ and $\theta _{2}$ (see Fig 3). One
may \emph{choose} to have a single number represent $\theta _{1},\theta _{2}$
and $\theta _{3}.$ A popular choice is the mean, which is calculated for
each marginal (see appendix for details),%
\begin{equation}
\left\langle \theta _{1}\right\rangle =0.2942,~\left\langle \theta
_{2}\right\rangle =0.1115,~\left\langle \theta _{3}\right\rangle =0.5942
\end{equation}

\begin{figure}[!t]
\center
\resizebox{5.5in}{2.5in} 
{\includegraphics[draft=false]{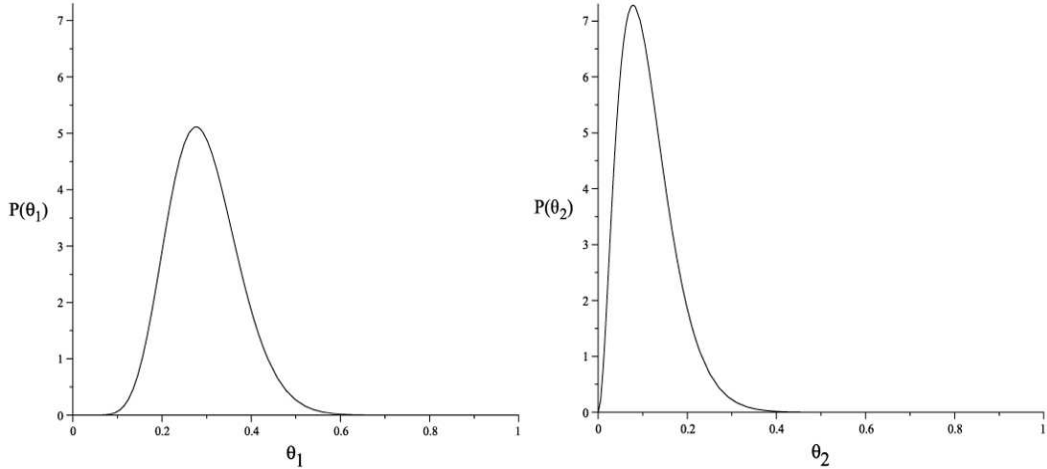}}
\caption{These figures show the distributions of $\protect\theta _{1}$ and $%
\protect\theta _{2}$ respectively.}
\label{Fig3:a}
\end{figure}

We now use the GJ solution (\ref{Econ Sanov Solution}) to compute the
"probabilities". We use the frequencies, $m/n$ for $Q$ or $Q_{1}=11/20,$ $%
Q_{2}=2/20$ and $Q_{3}=7/20$ and \emph{assume} that $F$ represents the
sample average for the entire population of balls. This produces the
following results:%
\begin{equation}
P_{1}^{\ast }=0.3015,~P_{2}^{\ast }=0.0971,~P_{3}^{\ast }=0.6015_{.}
\end{equation}%
Clearly the results are very close, however, there are several drawbacks to
using the Sanov approach. The first is that $P^{\ast }$ is estimated on the
basis of a frequency, $Q$ that is being used to represent an estimate of the
entire population. As is well known this can only be the case when $%
n\rightarrow \infty .$ MrE needs not make such assumptions. Similarly MrE
can incorporate actual expectation values, not sample averages disguised as
them. Second, the correct distribution to be used is the multinomial when
one is counting, not the frequencies of the observables. Third, and
practically most important, because the MrE solution produces a probability
distribution, one can take into account fluctuations. A single number would
not give any indication as to the uncertainty of the estimate. With our
method, one has the choice of which estimator one would like to use. Perhaps
the distribution is almost flat. Then our method would indicate that almost
any choice is equally likely. There is an underlying theme here:
probabilities are not equivalent to frequencies \emph{except} in the
asymptotic case. Therefore, if one wishes to know the probable outcome of a
problem in all cases, use MrE.

\section{Conclusions}

\label{Conclusions}

The realization that the MrE method incorporates MaxEnt and Bayes' rule as
special cases has allowed us to go beyond Bayes' rule and MaxEnt methods to
process both data and expected value constraints simultaneously. Therefore,
we would like to emphasize that anything one can do with Bayesian or MaxEnt
methods, one can now do with MrE. Additionally, in MrE one now has the
ability to apply additional information that Bayesian or MaxEnt methods
could not. Further, any work done with Bayesian techniques can be
implemented into the MrE method directly through the joint prior.

It is not uncommon to claim that the non-commutability of constraints
represents a \emph{problem} for the MrE method. Processing constraints in
different orders might lead to different inferences. We have argued that on
the contrary, the information conveyed by a particular sequence of
constraints is not the same information conveyed by the same constraints in
different order. Since different informational states should in general lead
to different inferences, the way MrE processes non-commuting constraints
should not be regarded as a \emph{shortcoming} but rather as a \emph{feature}
of the method.

Two specific econometric examples were solved in detail to illustrate the
application of the method. These cases can be used as templates for real
world problems. Numerical results were obtained to illustrate explicitly how
the method compares to other methods that are currently employed. The MrE\
method was shown to be superior in that it did not need to make asymptotic
assumptions to function and allows for fluctuations.

It must be emphasized that in the asymptotic limit, the MrE form is
analogous to Sanov's theorem. However, this is only one special case. The
MrE method is more robust in that it can also be used to solve traditional
Bayesian problems. In fact it was shown that if there is no moment
constraint one recovers Bayes rule.

\noindent \textbf{Acknowledgements:} I would like to acknowledge valuable
discussions with A. Caticha, M. Grendar, C. Rodr\'{\i}guez and E. Scalas.

\appendix

\section{Solving the normalization factor}

\label{Appendix}

Here we show how the means $\left\langle \theta _{1}\right\rangle
,\left\langle \theta _{2}\right\rangle $ and $\left\langle \theta
_{3}\right\rangle $ were calculated explicitly in the numerical solutions
section. The program Maple was used to calculate all results after the
integral from was created.

In general, we rewrite the posterior (\ref{posterior b}) in more detail,
dropping the superscripts,

\begin{equation}
P_{\text{new}}(\theta )=\frac{1}{\zeta ^{\prime }}\delta
(\sum\limits_{i}^{k}\theta _{i}-1)\prod\limits_{i=1}^{k}e^{\beta f_{i}\theta
_{i}}\theta _{i}^{m_{i}^{\prime }}.  \label{posterior c}
\end{equation}%
where $\zeta ^{\prime }$ differs from $\zeta $ in (\ref{zeta b}) only by a
combinatorial coefficient, 
\begin{equation}
\zeta ^{\prime }=\int \delta (\sum\limits_{i}^{k}\theta
_{i}-1)\prod\limits_{i=1}^{k}d\theta _{i}e^{\beta f_{i}\theta _{i}}\theta
_{i}^{m_{i}^{\prime }}~.  \label{zeta c}
\end{equation}%
A brute force calculation gives $\zeta ^{\prime }$ as a nested
hypergeometric series, 
\begin{equation}
\zeta ^{\prime }=e^{\beta f_{k}}I_{1}(I_{2}(\ldots (I_{k-1})))\,,  \label{Zc}
\end{equation}%
where each $I$ is written as a sum of $\Gamma $ functions,

\begin{equation}
I_{j}=\Gamma (b_{j}-a_{j})\sum\limits_{q_{j}=0}^{\infty }\frac{\Gamma
(a_{j}+q_{j})}{\Gamma (b_{j}+q_{j})~q_{j}!}t_{j}^{q_{j}}I_{j+1}~,
\end{equation}%
where $I_{k}=1.$ The index $j$ takes all values from $1$ to $k-1$ and the
other symbols are defined as follows: $t_{j}=\beta \left(
f_{k-j}-f_{k}\right) $, $a_{j}=m_{k-j}^{\prime }+1$ and%
\begin{equation}
b_{j}=n+j+1+\sum\limits_{i=0}^{j-1}q_{i}-\sum\limits_{i=0}^{k-j-1}m_{i}^{%
\prime }~\,,
\end{equation}%
with $q_{0}=m_{0}^{\prime }=0$. The terms that have indices $=0$ are equal
to zero (i.e. $b_{0}=q_{0}=0,$ etc.). A few technical details are worth
mentioning: First, one can have singular points when $t_{j}=0$. In these
cases the sum must be evaluated in the limit as $t_{j}\rightarrow 0.$
Second, since $a_{j}$ and $b_{j}$ are positive integers the gamma functions
involve no singularities. Lastly, the sums converge because $a_{j}>b_{j}$.
The normalization for the first example (\ref{Z2}) can be calculated in a
similar way.

Specifically for (\ref{MrE Posterior}), the Lagrange multiplier $\beta $ was
determined using Newton's method on the equation (\ref{F b}) and found to be 
$\beta =14.1166.$ This result is then put into (\ref{Zc}) in order to attain 
$\zeta ^{\prime }=1874.1247.$ Next, the means were calculated by increasing $%
m_{i}+1$ and $n+1,$ then recalculating so that 
\begin{equation}
\begin{array}{l}
\left\langle \theta _{1}\right\rangle =\frac{\zeta _{m_{1}+1,n+1}^{\prime }}{%
\zeta ^{\prime }}~, \\ 
\left\langle \theta _{2}\right\rangle =\frac{\zeta _{m_{2}+1,n+1}^{\prime }}{%
\zeta ^{\prime }}~, \\ 
\left\langle \theta _{3}\right\rangle =1-\left\langle \theta
_{1}\right\rangle -\left\langle \theta _{2}\right\rangle ~.%
\end{array}%
\end{equation}

Currently, for small values of $k$ (less than 10, depending on memory) it is
feasible to evaluate the nested sums numerically; for larger values of $k$
it is best to evaluate the integral for $\zeta ^{\prime }$ using sampling
methods.

\end{document}